\documentclass[twocolumn,floatfix,showpacs,prl,aps]{revtex4}

\usepackage[matrix,frame,arrow]{xypic}
\usepackage{amsfonts,amsmath,amssymb}
\usepackage{graphicx,epsfig} 

\input{Qcircuit}

\newcommand{\bs}{$\mathcal{C}_{BS}^{(n)}$ }
\newcommand{\bsc}[1]{$\mathcal{C}_{BS}^{(#1)}$}

\begin{document}

\author{Panos Aliferis$\,^{1,3}$}

\author{Andrew W. Cross$\,^{2,3}$}

\title{Subsystem fault tolerance with the Bacon-Shor code}

\affiliation{\vspace*{1.2ex}
             $^1$ {Institute for Quantum Information, California Institute of Technology, Pasadena, CA 91125}\\
             $^2$ {Massachusetts Institute of Technology, 77 Massachusetts Avenue, Cambridge, MA 02139} \\
	     $^3$ {IBM Research Division, T. J. Watson Research Center, P.~O. Box 218, Yorktown Heights, NY 10598}}

\date{\today}

\begin{abstract}


We discuss how the presence of gauge subsystems in the Bacon-Shor code [D.~Bacon, {\it Phys. Rev.} A {\bf 73}, 012340 (2006)] leads to remarkably simple and efficient methods for fault-tolerant error correction (FTEC). Most notably, FTEC does not require entangled ancillary states and it can be implemented with nearest-neighbor two-qubit measurements. By using these methods, we  prove a lower bound on the quantum accuracy threshold, $1.94{\times}10^{-4}$ for adversarial stochastic noise, that improves previous lower bounds by nearly an order of magnitude. 



\end{abstract}

\maketitle
 
 
Operating a full-scale quantum computer will require methods for protecting against decoherence or systematic hardware imperfections. 
One of the central results in the theory of fault-tolerant quantum computation \cite{Shor96}, the quantum threshold theorem shows that a noisy quantum computer can accurately and efficiently simulate any ideal quantum computation provided that noise is {\em local} and its strength is below a critical value known as the quantum {\em accuracy threshold} \cite{note4}. 
The actual value of the accuracy threshold is therefore of great practical interest as it represents the desired target accuracy of prospective implementations of quantum computation. However, determining this value is in general a very challenging problem since not only does it depend on the particular character of noise but it also depends on the specifics of the quantum computing architecture (e.g., geometric constraints on qubit interactions). In this Letter, we present methods of fault-tolerant quantum computation based on a new subsystem code that significantly improve the existing lower bounds on the value of the quantum accuracy threshold 
and that are also beneficial for geometrically local quantum computing architectures.

Subsystem codes protect quantum information from noise by mapping the system to be encoded (e.g., one logical qubit) into a {\em subsystem} of a larger system \cite{Knill99,Kribs05}; 
in fact, such a mapping provides the most general possible encoding of quantum information \cite{Knill06}. 
More concretely, if we let $\mathcal{H}_S$ be the Hilbert space of the system to be encoded, the encoding maps density operators and observables from $\mathcal{H}_S$ to a subsystem with Hilbert space $\mathcal{H}_{L}$ which lies inside a larger Hilbert space, $\mathcal{H} = \left( \mathcal{H}_{L} \otimes \mathcal{H}_T \right) \oplus \mathcal{H}_R$, where $\mathcal{H}_T$ describes additional ``gauge'' subsystems not necessarily protected from noise and $\mathcal{H}_R$ labels the ``rest'' of $\mathcal{H}$.  Fault-tolerant quantum computation has traditionally used subspace codes that can be seen as subsystem codes where $\mathcal{H}_T$ is one-dimensional encoding {\em no} subsystem.   

Interestingly, the converse is also true: For any subsystem code there exists a corresponding subspace code that can be obtained by ``picking a gauge,'' i.e., by projecting the state in $\mathcal{H}_T$ onto a pure state \cite{Kribs05}. Because of this connection, subsystem codes do not have properties that make them fundamentally different from subspace codes. Nevertheless, subsystem codes are in a certain sense more efficient than the corresponding subspace codes since they require the extraction of fewer syndrome bits \cite{Poulin05,Bacon05} (but, see also \cite{Aly06}). As we will discuss in this Letter, subsystem codes are also advantageous for quantum {\em computation} because the presence of gauge subsystems can help us simplify the quantum circuits that implement fault-tolerant error correction.


We will consider fault-tolerant quantum circuits where computation is encoded using a subsystem code due to Bacon \cite{Bacon05}---because of the close relation of this code with Shor's code \cite{Shor95b}, we will refer to it as the {\em Bacon-Shor code}. There is a different Bacon-Shor code for every integer \parbox{0.9cm}{$n>1$}; for fixed $n$, the corresponding code, $\mathcal{C}_{BS}^{(n)}$, is a distance-$n$ stabilizer CSS code \cite{Gottesman97} encoding one protected logical qubit into a code block of $n^2$ physical qubits.

\begin{figure}[t]
\begin{center}
\epsfig{file=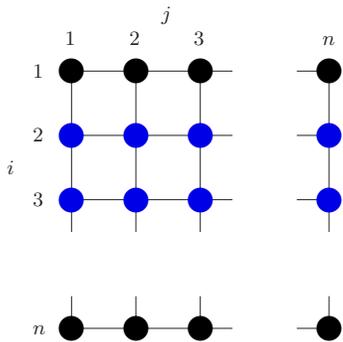,width=4.5cm} \vspace{-0.3cm}
\end{center}
\caption{\label{fig:1} 
Qubits in the \bs block sit on the vertices of an $n{\times}n$ square lattice. An element of the code stabilizer, the operator $X_{2,*}X_{3,*}$ applies $X$ on all qubits shown in blue.
         }
\end{figure}

If we imagine placing the $n^2$ qubits in the \bs block on the {\em vertices} of an $n{\times}n$ square lattice as in Fig.$\,$\ref{fig:1}, the code's stabilizer group \cite{Gottesman97}, $\mathcal{S}$, is generated by 
%
\begin{equation}
\label{eq:stab} 
\mathcal{S} = \langle X_{j,*}X_{j+1,*} \, ; \, Z_{*,j}Z_{*,j+1} \, | \, j\in \mathbb{Z}_{n-1} \rangle \, ,
\end{equation}  
\noindent where we have used the short-hand notation for the Pauli matrices $X\equiv \sigma_{\rm x}$, $Z\equiv \sigma_{\rm z}$, and $O_{j,*}$, $O_{*,j}$ denote operators that act nontrivially as a tensor product of $O$ operators on all qubits in row or column $j$, respectively. Throughout, $\mathbb{Z}_{m}$ is understood to indicate the set $\{1,2,\dots,m \}$.

The code {\em syndrome} $\vec{e}$---i.e., the binary vector of eigenvalues of the $2(n{-}1)$ stabilizer generators in Eq.$\,$(\ref{eq:stab})---induces a decomposition of the Hilbert space, $\mathcal{H}$, of the $n^2$ qubits in the code block into subspaces encoding $n^2-2(n{-}1)=(n{-}1)^2+1$ logical qubits. Therefore, within each subspace with fixed syndrome---and, in particular, within the {\em code space} corresponding to the trivial syndrome---we can define a subsystem decomposition
\begin{equation}
\label{eq:dec}
\mathcal{H}= \bigoplus_{\vec{e}} \left( \mathcal{H}_L \otimes \mathcal{H}_{\rm T} \right) \, \vspace{-0.1cm},
\end{equation}
\noindent where we associate $\mathcal{H}_L$ with the one logical qubit protected by the full distance, $n$, of the code. The logical Pauli operators for this logical qubit can be defined as $X_L=X_{1,*}$ (i.e., a tensor product of $X$ operators applied on all qubits in the first row) and $Z_L=Z_{*,1}$ (i.e., a tensor product of $Z$ operators applied on all qubits in the first column).
The remaining $(n{-}1)^2$ logical qubits live in $\mathcal{H}_T$ and their logical Pauli operators can be chosen from the non-abelian group
\begin{equation}
\label{t}
 T = \langle \; X_{j,i} X_{j+1,i} \; ; Z_{i,j} Z_{i,j+1} \; | \; i\in \mathbb{Z}_{n} \; ; j\in \mathbb{Z}_{n{-}1} \;  \rangle \; ,
\end{equation}
\noindent where we have used the notation $O_{i,j}$ for an operator $O$ acting on the qubit with coordinates $(i,j)$. Indeed, the operators in $T$ commute with every operator in the code stabilizer, they commute with the logical operators $X_L$ and $Z_L$ and, furthermore, they can be grouped into $(n{-}1)^2$ independent pairs of anticommuting operators  with operators in different pairs commuting. 

Given some nontrivial syndrome value, error recovery for the logical qubit encoded in $\mathcal{H}_{L}$ proceeds in a similar manner as in the classical repetition code: The eigenvalues of the stabilizer generators $\{X_{j,*}X_{j+1,*}\}$ can be used to correct $Z$ errors on up to $\lfloor n/2 \rfloor$ rows. Moreover, only the parity of $Z$ errors in each row is relevant: an operator acting as $Z$ on a pair of qubits at the same row is an operator in $T$ and, therefore, has no effect on the protected information in $\mathcal{H}_L$. Error recovery for $X$ errors proceeds similarly along the columns. 


Since the logical Pauli operators for the logical qubits encoded in $\mathcal{H}_T$ act nontrivially on only two qubits in the code block, if we take into consideration all $(n{-}1)^2+1$ logical qubits then \bs has distance 2 and it is an error {\em detecting} code. However, if we only consider the logical qubit encoded in $\mathcal{H}_{L}$ then the effective distance is $n$ and errors with support on up to $\lfloor n/2 \rfloor$ qubits in the code block can be corrected---we will call this logical qubit the {\em protected} qubit. In fact, error recovery for the protected qubit may unavoidably result in applying at the same time nontrivial logical operations with support on $\mathcal{H}_{T}$. This is not a problem as long as we never encode quantum information in $\mathcal{H}_{T}$. We can think of the $(n{-}1)^2$ logical qubits encoded in $\mathcal{H}_{T}$ as {\em gauge} qubits since they correspond to degrees of freedom for the logical information encoded in $\mathcal{H}_{L}$. In some cases it will be sufficient to completely disregard the state of the gauge qubits. More interestingly, we will next discuss how fault-tolerant error correction for the protected qubit can be simplified by taking advantage of the presence of gauge qubits.

\begin{figure}[ht]
\begin{center}
\begin{tabular}{ccc}
\put(-0.8,0.1){(a)}
\put(3.1,0.1){(b)}
\Qcircuit @C=1ex @R=2.3ex @!R {  \put(0.1,0.4){\footnotesize{$j,k$}}
   & \qw                             & \targ      & \qw       & \qw     & \qw \\
   & \qw  \put(-0.4,0.4){\footnotesize{$j{+}1,k$}}  & \qw        & \targ     & \qw     & \qw \\
   & \push{|+\rangle \hspace{0.1cm}} & \ctrl{-2}  & \ctrl{-1} & \meterx &
                                }    
   &\hspace{1.2cm} & \hspace{0.2cm}
\Qcircuit @C=1ex @R=2.3ex @!R {  \put(0.1,0.4){\footnotesize{$k,j$}}
   & \qw                             & \ctrl{+2}  & \qw       & \qw     & \qw \\
   & \qw  \put(-0.4,0.4){\footnotesize{$k,j{+}1$}}  & \qw        & \ctrl{+1} & \qw     & \qw \\
   & \push{|0\rangle \hspace{0.1cm}} & \targ      & \targ     & \meterz &
                                }    
 \vspace{-0.2cm}
\end{tabular}
\end{center}
\caption{\label{fig:2} (a) A circuit for measuring the operator $X_{j,k}X_{j+1,k}$ using one ancillary qubit. (b) A similar circuit for measuring $Z_{k,j}Z_{k,j+1}$. $|+\rangle \propto |0\rangle + |1\rangle$ is the $+1$ eigenstate of $X$.}
\end{figure}
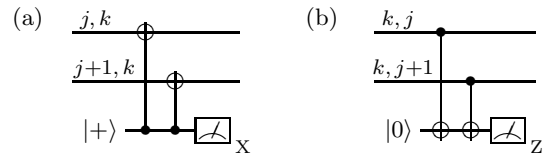

Let us first explain how we can extract the code syndrome indirectly by manipulating the state of the gauge qubits. Our first observation is that we can express the stabilizer generators in Eq.$\,$(\ref{eq:stab}) as 
\begin{equation}
\begin{array}{rcl}
\label{eq:gauge}
X_{j,*}X_{j+1,*} & = & \bigotimes _{k=1}^n \left( X_{j,k} X_{j+1,k} \right) \; ; \vspace{0.1cm}\\
Z_{*,j}Z_{*,j+1} & = & \bigotimes _{k=1}^n \left( Z_{k,j}\; Z_{k,j+1} \right) \; .
\end{array}
\end{equation}  
\noindent What is remarkable about this decomposition is that the operators in parentheses are supported on $\mathcal{H}_T$; 
hence, they commute with operators in the code stabilizer and also commute with the logical operators for the protected qubit. Because of this, we can \emph{measure} each of them \emph{separately} and then Eq.$\,$(\ref{eq:gauge}) implies that the code syndrome can be computed by taking the appropriate parities of the measurement outcomes. Moreover, since these operators act nontrivially on only two qubits in the code block, measuring them is especially easy; e.g., Fig.$\,$\ref{fig:2} shows simple circuits for measuring $X_{j,k}X_{j+1,k}$ and $Z_{k,j}Z_{k,j+1}$. 

%
%

This indirect method for inferring the code syndrome 
significantly reduces the overhead for fault-tolerant error correction (FTEC). This is because, unlike in standard FTEC methods \cite{note4}, this method does not require preparing and verifying entangled ancillary states---ancillary qubits in the $|0\rangle$ or $|+\rangle$ state are sufficient as in Fig.$\,$\ref{fig:2}. Furthermore, for maximum qubit efficiency but at a cost of extra memory error, even a {\em single} ancillary qubit suffices to sequentially measure all two-qubit operators necessary to extract the syndrome! In addition, the specific two-qubit operators to be measured can be chosen to have support on neighboring qubits in the code block when these qubits are arranged on a two-dimensional square lattice. For this reason, this FTEC method could prove to be especially advantageous for quantum computing with geometric locality constraints. 


In settings where nonlocal qubit interactions are possible or when qubit movement has error rate much lower than quantum gates, other standard FTEC methods will yield the best accuracy thresholds. Since \bs is a CSS code, Steane's FTEC method \cite{Steane96b} can be used to extract the syndrome provided we can fault-tolerantly prepare logical $|0\rangle$ ($|0\rangle_L$) and logical $|+\rangle$ ($|+\rangle_L$) states for the protected qubit. Alternatively, Knill's FTEC method \cite{Knill05} requires fault-tolerantly preparing two code blocks where the two protected qubits are encoded in a logical Bell state ($|\Phi_0\rangle_L \propto |0\rangle_L|0\rangle_L + |1\rangle_L|1\rangle_L$). 
 
As is evident from the decomposition in Eq.$\,$(\ref{eq:dec}), the distinctive feature of subsystem codes is that logical states in $\mathcal{H}_L$ are not uniquely encoded: after having specified a logical state in $\mathcal{H}_L$, the state in $\mathcal{H}_T$ can still be arbitrary. This freedom in choosing the state of the gauge qubits can be used to our advantage in the design of encoding circuits of logical states for the protected qubits. In particular, we will next discuss how we can design remarkably simple encoding circuits for the logical ancillary states required for Steane's and Knill's FTEC methods, thus also reducing the overhead associated with  the post-encoding verification of these states.

For concreteness, consider designing an encoding circuit for $|0\rangle_L$, i.e., a state in the code space which is the ${+}1$ eigenstate of $Z_L$. With the state in $\mathcal{H}_L$ specified, we can choose the  state in $\mathcal{H}_T$ to be the ${+}1$ eigenstate of the gauge-qubit operators $\{X_{i,j} X_{i+1,j}\}$. In other words, our encoding circuit prepares the $+1$ eigenstate of the operators in the following stabilizer group:
\begin{equation}
\label{0-stab}
\mathcal{S}(|0\rangle_L) = \langle X_{i,j}X_{i+1,j} \; ; Z_{*,j} \; | \; i\in\mathbb{Z}_{n{-}1} ; j\in\mathbb{Z}_{n} \rangle \; .
\end{equation}
\noindent We recognize the state described by Eq.$\,$(\ref{0-stab}) as a tensor product of $n$ Schr$\ddot{\rm o}$dinger ``cat'' states in the Hadamard-rotated basis, each one lying across a column in Fig.$\,$\ref{fig:1}.  
   
We can obtain the state $|+\rangle_L$ by applying a logical Hadamard transformation to the state $|0\rangle_L$. We observe that applying Hadamard gates {\em bitwise} has the same effect as a logical Hadamard gate up to a 90-degree rotation of the square lattice in Fig.$\,$\ref{fig:1} (and, also, up to a nontrivial operation acting on the gauge qubits). We can therefore obtain $|+\rangle_L$ by first preparing $|0\rangle_L$, applying Hadamard gates bitwise and, finally, rotating the lattice in Fig.$\,$\ref{fig:1} by 90 degrees. The bitwise Hadamard gates will produce usual cat states in the computation basis which, after rotating the lattice by 90 degrees, will be aligned each to lie across a row---this is our construction for $|+\rangle_L$. In addition, logical Bell states for Knill's FTEC method can be constructed by interacting two blocks encoded in the states $|+\rangle_L$ and $|0\rangle_L$ via a logical {\sc cnot} gate. Since the logical {\sc cnot} gate can be implemented by transversal {\sc cnot} gates, this construction is also especially simple. 

The logical {\sc cnot} and Hadamard gates have simple transversal implementations and, together with single-qubit preparation and measurement, they suffice for performing fault-tolerant error correction; however, they are not universal for quantum computation. 
Universal quantum computation can be realized by including the logical phase gate, $S\equiv \exp(-i {\pi \over 4} \sigma_{\rm z} )$, and one logical non-Clifford gate in our gate set; for a detailed discussion on achieving encoded quantum universality see chapter 5 in \cite{sup}.

We have analysed the accuracy threshold of fault-tolerant quantum circuit simulations that use the concatenated \bs for various values of $n$. In our analysis, we have considered {\em adversarial stochastic noise} \cite{Aliferis05b}, i.e., the stochastic form of {\em local noise} \cite{TB05,Aliferis05b,sup}. In this noise model, we assume that any $r$ {\em specific} elementary physical operations in the noisy quantum circuit (single-qubit preparations, quantum gates, memory steps or single-qubit measurements) fail with probability at most $\varepsilon^r$ for some constant $0\leq \varepsilon \leq 1$. Noise is adversarial because faults need not be independent and, moreover, noise may act jointly on the support of all different faulty operations. 
Adversarial stochastic noise is the natural form of noise to consider when analyzing fault-tolerant circuit simulations that use concatenated codes. This is because even if physical noise does not include adversarial correlations between different faulty locations, the {\em effective} noise that acts on all coding levels of the concatenation hierarchy higher than the physical level may include such correlations which arise due to coding and the propagation of syndrome information \cite{Aliferis05b}. 

Our rigorous lower bounds on the quantum accuracy threshold were obtained by performing an analysis of malignant sets of locations on extended rectangles according to the method that was introduced in \cite{Aliferis05b}. 
%
%
Table \ref{table:1} summarizes our results. We carried out the required combinatorial analysis for the concatenated \bsc{3} using both Steane's and Knill's FTEC methods and for the concatenated \bsc{5} using only the former method due to time limitations. The analysis was done by using a computer program running for several months on a cluster of 20 Pentium III processors. Our best rigorous lower bound on the accuracy threshold, $1.94\times 10^{-4}$, was obtained for the concatenated \bsc{5}. This lower bound improves by nearly an order of magnitude the $2.73{\times}10^{-5}$ lower bound that was proved in \cite{Aliferis05b} with the concatenated Steane [[7,1,3]] code.

\begin{table}[t]
\begin{tabular}{r|c|c|c|c}
  Code $\;$ & $\,$FTEC$\,$ & $\,$locs.$\,$ & $\,\varepsilon_0\,(\times 10^{-4})$  &  $\varepsilon^{\rm MC}_0\,(\times 10^{-4})$ \\ 
Steane [[7,1,3]] $\;$ &  Steane     &  $\;$ 575  $\;$                 & $\,0.27\,$ & \\
\bsc{3} [[9,1,3]] $\;$ &  Steane     &  $\;$ 297 $\;$                  & $\,1.21\,$     & $1.21\pm 0.06$ \\ 
        $\;$  &  Knill     &   297                   & $\,1.26\,$     & $1.26\pm 0.05$ \\ 
\bsc{5} [[25,1,5]] $\;$ &  Steane     & $\;$ 1,185 $\;$                  & $\bf \,1.94\,$ & $1.92\pm 0.02$ \\ 
        $\;$  &  Knill     & 1,185                   &                          & $\bf\,2.07\pm 0.03$ \\ 
Golay [[23,1,7]] $\;$ &  Steane     & $\;$ 7,551  $\;$                 &                          & $\approx 1$  \\ 
\bsc{7} [[49,1,7]] $\;$ &  Steane     & $\;$ 2,681 $\;$                 &                          & $\,1.74\pm 0.01$ \\ 
        $\;$ &  Knill     & $\;$ 2,681 $\;$                  &                          & $\,1.91\pm 0.01$ \\ 
\end{tabular} 
\caption{\label{table:1} Rigorous lower bounds on the accuracy threshold, $\varepsilon_0$, for adversarial stochastic noise  with the concatenated Bacon-Shor code of varying block size and comparison with prior rigorous lower bounds using the concatenated Steane [[7,1,3]] code \cite{Aliferis05b} and Golay [[23,1,7]] code \cite{RO}. The third column gives the number of locations in the {\sc cnot} extended rectangle \cite{Aliferis05b}. The forth column gives exact lower bounds on $\varepsilon_0$; the results are obtained using a computer-assisted combinatorial analysis. The fifth column is the Monte-Carlo estimate for $\varepsilon_0$ with $1\sigma$ uncertainties. 
Bold fonts indicate the best results in each column.
}
\end{table}

Analyzing codes of larger block size than \bsc{5} proved to be computationally difficult in this exact setting. In these cases, we have used a Monte-Carlo method to uniformly sample the set of fault paths with a fixed number of faulty locations inside an extended rectangle. By estimating what fraction, $\hat{f}$, of these sets is malignant, we obtain an estimate of the exact combinatorial coefficients that determine the accuracy threshold with a standard error that can be determined by using $\sigma_{\hat{f}}=\sqrt{\hat{f}(1-\hat{f})/N}$ where $N$ is the sample size. We have also applied this Monte-Carlo method to the cases where we could extract the exact combinatorics in order to provide evidence that the Monte-Carlo estimates are accurate. Indeed, as can be seen in Table \ref{table:1}, the exact lower bounds in those cases lie within $1\sigma$ of the estimated lower bounds. Our Monte-Carlo results give evidence that the quantum accuracy threshold achieves a maximum over all concatenated Bacon-Shor codes of varying block size for the 25-qubit code using Knill's FTEC method. Chapter 5 in \cite{sup} discusses the details of our threshold analysis for the {\sc cnot} extended rectangle and explains why the accuracy threshold is determined by this gate and not by non-Clifford gates.


In conclusion, we have shown how the presence of gauge qubits in the Bacon-Shor code can be exploited to design quantum circuits for fault-tolerant error correction with remarkable properties. We have presented a new method for fault-tolerant error correction that uses nearest-neighbor two-qubit measurements and does not require the preparation of entangled ancillary states. We expect this method to be advantageous in implementations that impose geometric locality constraints in the interactions between qubits such as, e.g., in proposed ion-trap or solid-state schemes of quantum computation. On the other hand, standard methods for fault-tolerant error correction can be implemented using especially simple encoding circuits for the required ancillary logical states, thus greatly reducing the overhead associated with the verification of these states. Our lower bound on the quantum accuracy threshold, $1.94{\times}10^{-4}$ for adversarial stochastic noise, is the best that has been rigorously proven so far and improves previous rigorous lower bounds by nearly an order of magnitude. 

An open challenge for future work is to investigate to what extent the lower bounds we have presented can be improved by performing a more detailed combinatorial analysis or by considering a less adversarial noise model. In addition, it would be important to derive lower bounds on the accuracy threshold in a setting where qubit interactions are geometrically local in order to quantify the advantages of our FTEC methods and to compare with other fault-tolerant schemes for local quantum computation such as, e.g., those in \cite{Svore06,RH06}. 


We are grateful to Ike Chuang, David DiVincenzo, Debbie Leung, John Preskill, Krysta Svore, and Barbara Terhal for helpful discussions and suggestions. AC thanks John Preskill for the invitation to the Caltech Institute for Quantum Information, where some of this work was done. PA is supported by the NSF under grant no.$\,$PHY-0456720. AC is partially supported by a research internship at IBM T. J. Watson Research Center.

\bibliographystyle{unsrt}

\end{document}